\documentclass[twocolumn,amsmath,amssymb,aps,pre,showpacs]{revtex4}
\usepackage{graphics}

\begin{document}

\title{Fluctuation formula and non-Gaussian distribution 
       in the thermostated Lorentz gas}

\author{M. Dolowschi{\'a}k} \email{dolowsch@szerenke.elte.hu} 
\author{Z. Kov{\'a}cs} \email{kz@garfield.elte.hu}

\affiliation{Institute for Theoretical Physics, E{\"o}tv{\"o}s University, 
          Pf.\ 32, H--1518 Budapest, Hungary}

\begin{abstract}
In this paper we numerically examine the connection of the Gallavotti-Cohen 
fluctuation formula and the functional form of the corresponding probability 
density function in the field driven Lorentz gas  thermostated by the Gaussian
isokinetic thermostat. 
We analyze the moments of the entropy production rate fluctuations 
and show that all the central moments of the averaged fluctuations exhibit 
{\em power law} dependence on the length of the averaging time interval,
indicating that this density deviates from a Gaussian. 
Furthermore the obtained exponents are found to obey a special 
{\em pairing rule} showing that the corresponding probability density 
function can not be scaled in the averaging time interval.
\end{abstract}

\pacs{05.45.-a, 05.70.Ln}

\maketitle

\section{Introduction}

In recent years a large number of papers have focused on the  
Gallavotti-Cohen fluctuation formula (FF) with
both theoretical and numerical tools in various nonequilibrium systems
\cite{shearflow,dynens,onsext,stocha,stocha2,exptest,stdystat,maes,
flin2d,tamas}.
The FF is a symmetry property of the probability density function (PDF) of 
a dynamically measured quantity $\xi$ connecting the probabilities 
of measuring $\xi$ values with equal magnitudes but opposite signs.
This property was first observed numerically in a system of 
thermostated fluid particles undergoing shear flow \cite{shearflow}
where $\xi$ was the phase space contraction rate.
The significance of this result was largely affected by the fact that
the phase space contraction rate could be connected to the 
entropy production rate, thus the FF gave an insight to the
nonequilibrium thermodynamical behavior of the system. 
Since then the nontrivial connection of the phase space contraction rate and 
the entropy production rate has been investigated and was treated in 
\cite{epr}.

More precisely,
let $\xi_\tau(t)$ denote the quantity $\xi$ 
averaged over a time interval of length $\tau$ centered around time $t$:
$\xi_\tau(t)= \frac{1}{\tau}
      \int\limits_{-{\tau}/{2}}^{{\tau}/{2}}\xi(t+t')\,\mathrm{d}t'$. 
Considering it as a stochastic variable, 
its statistical properties in a steady state can be 
characterized by a probability density function $\Xi_\tau(x)$. 
The FF states that this PDF has the following  
property:
\begin{equation}
  \label{flform}
  \lim_{\tau\to\infty}
        \frac{1}{\tau} \ln \frac{\Xi_\tau(x)}{\Xi_\tau(-x)}=x\,. 
\end{equation}
In other words, the probability of observing a negative $\xi$ value
is exponentially smaller than that of the corresponding positive value
in the $\tau \to \infty$ limit.

The FF has also been found to be valid in certain systems with
strong external forcing \cite{chaodyn}, 
thus it is expected to say something important 
about systems far from equilibrium 
(as opposed to linear response theory,
which is valid for vanishing external fields).
Consequently, significant theoretical efforts have been made 
to find a common property behind this behavior in all the systems obeying 
the FF \cite{dynens,stocha,stocha2,maes}.

Since the FF connects the values of the $\Xi_\tau(x)$ function 
on the negative half axis to the values of $\Xi_\tau(x)$ on the positive one,
but it does not state anything about the functional form of $\Xi_\tau(x)$,
it is interesting to examine the latter in systems found to obey the FF.   
This question is emphasized by the observation that 
Eq.~(\ref{flform}) can be satisfied by a Gaussian PDF
with a subcondition connecting its average and variance. 
This raises the necessity to study how close the underlying 
PDF of $\Xi_\tau(x)$ is to a Gaussian in systems which seem to obey the FF.

In a previous paper \cite{flform} we investigated numerically 
the validity of the FF in the two and three 
dimensional periodic LG subject to a constant electric and 
magnetic field and thermostated by the GIK thermostat. 
We found that the symmetry property described by the FF is
valid for low and high $\tau$ values in the reversible configurations 
and seems to be approximately valid in the $\tau \to \infty$ limit 
even in the {\em absence} of reversibility.
This also motivated us to study the $\tau$-dependence of the details
of $\Xi_\tau(x)$ in the thermostated LG.  

In this paper we numerically analyze the connection of obeying the FF
and the details of the functional form of the corresponding PDF
in the periodic Lorentz Gas (LG) thermostated by a Gaussian 
Isokinetic (GIK) thermostat.   
In Sec.~\ref{sec:sys} we describe the simulated system, 
in Sec.~\ref{sec:result} we present our numerical results and in 
Sec.~\ref{sec:concl} we summarize our conclusions.

\section{The system}
\label{sec:sys}

One of the simplest and most investigated models suitable for studying 
transport phenomena is the field driven Lorentz gas (LG) 
thermostated by a Gaussian isokinetic (GIK) thermostat.
This model consists of a charged particle subjected to 
an electric field moving in the lattice of elastic scatterers.
Due to the applied electric field, one must use a thermostating mechanism 
to achieve a steady state in the system.
Such a tool is the Gaussian isokinetic thermostat which preserves 
the kinetic energy of the particle \cite{thermostats}.

For the sake of simplicity, we investigate in our study the two dimensional (2D) 
periodic Lorentz gas with circular scatterers arranged on a square lattice.
We present the equations of motion in dimensionless variables:
choosing the unit of mass $m$ and electric charge $q$ to be equal to 
the mass and electric charge of the particle, yields $m=q=1$ in our 
model.
The unit of distance is taken to be equal to the radius of scatterers ($R=1$),
and the unit of time is chosen to normalize the magnitude of particle
velocity to unity.
Let ${\bf q}=\left(q_1,q_2\right)$ denote the position and
${\bf p}=\left(p_1,p_2\right)$ the momentum 
of the particle ($\left|{\bf p}\right|=1$).
The phase space variable of the system is 
${\mathbf \Gamma}=\left({\bf q},{\bf p}\right)$;
it is transformed abruptly at every elastic collision and evolved 
smoothly by the differential equation
\begin{eqnarray}
  \label{diffeq}
   \dot{\bf q} & = & {\bf p}       \nonumber \\
   \dot{\bf p} & = & {\bf E} - \alpha {\bf p}
\end{eqnarray}
between the collisions. 
Here $\alpha$  is called the {\em thermostat variable}, 
while ${\bf E}$ is the constant vector playing the role 
of the external electric field.
The GIK thermostat requires the fixing of the kinetic energy, which
leads to the choice $\alpha={\bf E}{\bf p}$ in Eq.~(\ref{diffeq}).


Dissipation in such models can be measured by the 
phase space contraction rate $\sigma$.
It can be computed by taking the divergence of the right-hand 
side of Eq.~(\ref{diffeq}):
\begin{eqnarray}
  \label{sigma}
   \sigma= - \mbox{div}\, {\bf \dot\Gamma} =\alpha\,,
\end{eqnarray}
and it can be shown that in this model
\begin{eqnarray}
\label{sigmaxi}
       \sigma(t)&=&\xi(t) ,
\end{eqnarray}
where $\xi$ is the entropy production rate defined according to the
$\xi=\frac{\bf JE}{k_B T}$ expression of irreversible thermodynamics, 
and the temperature $T$ is given by analogy with kinetic theory \cite{epr}.  
We note that while Eq.~(\ref{sigmaxi}) is valid for the GIK thermostat, 
it need not be true in general, as shown e.g.\ in models 
thermostated by some variants of the Nos{\'e}-Hoover thermostat \cite{nosehoover}
or by deterministic scatterings \cite{detscatt}.

\section{Numerical results}
\label{sec:result}

\begin{figure}
\scalebox{0.3}{\rotatebox{-90}{\includegraphics{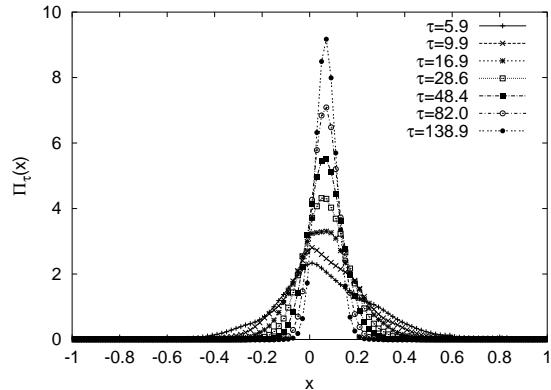}}}
\caption{The probability density $\Pi_\tau(x)$ in the 2D  
periodic LG, with lattice constant $d=2.1$
and external field ${\bf E}=(0.3,0.48)$,
the number of collisions is $1.58\times10^{7}$ and the averaged time between
two collisions is  $\approx 0.62$. 
For higher $\tau$ values the curves seems to be ``Gaussian like'': 
fitting a Gaussian function onto the measured values yields very good 
visual agreement.}
\label{histfig}
\end{figure}

The goal of our numerical simulations is to measure $\Xi_\tau(x)$
with a precision that is sufficient to check the validity of the fluctuation
formula, as well as the question of Gaussianity of the distribution.
Due to Eq.~(\ref{sigmaxi}), $\Xi_\tau(x)$ can be measured by 
periodically computing 
$\xi_\tau$ along a long particle trajectory and making a histogram of 
these data. 
The disadvantage of this method is that the range of possible
$\xi_\tau$ values depends on the strength of the electric field.
Instead of $\xi_\tau$, we may introduce the quantity
\begin{eqnarray}
  \label{pitau}
  \pi_\tau(t)= \frac{1}{\tau}
 \int\limits_{-\frac{\tau}{2}}^{\frac{\tau}{2}} 
               {\bf n_E}{\bf p}(t+t^{'})\,\mathrm{d}t^{'},   
\end{eqnarray}
where ${\bf n_E}$ denotes the unit vector parallel to ${\bf E}$. 
Since the magnitude of ${\bf p}$ is unity, $\pi_\tau$ always satisfies 
$\pi_\tau\in[-1,1]$.
By making a histogram of the periodically measured values of $\pi_\tau$, 
one gets an approximation of its probability density $\Pi_\tau(x)$ shown 
in Fig.~\ref{histfig}.
The two stochastic variables defined above satisfy $\xi_\tau=E\pi_\tau$, 
so the relationship between two probability densities $\Xi_\tau$ and 
$\Pi_\tau$ is simply
$
  \label{XiPi}
   \Xi_\tau(x)=\frac{1}{E}\,\Pi_\tau\left(\frac{x}{E}\right).
$

\subsection{Scaling}
We can observe in Fig.~\ref{histfig} that for high enough $\tau$ values 
the PDF $\Pi_\tau(x)$ seems to be close to a Gaussian. 
In order to examine this behavior we applied 
the following scaling transformation to the $\Pi_\tau(x)$ functions:
\begin{eqnarray}
\label{eqscal}
S_\tau(x)=\frac{1}{\Pi^{\left(max\right)}_\tau}
\Pi_\tau\left(\frac{x}{\Pi^{\left(max\right)}_\tau}+x_{max}\right).
\end{eqnarray}
In this equation $x_{max}$ denotes the $x$-value where $\Pi_\tau(x)$ 
takes  its maximal value denoted by $\Pi^{\left(max\right)}_\tau$.
\begin{figure}
\scalebox{0.3}{\rotatebox{-90}{\includegraphics{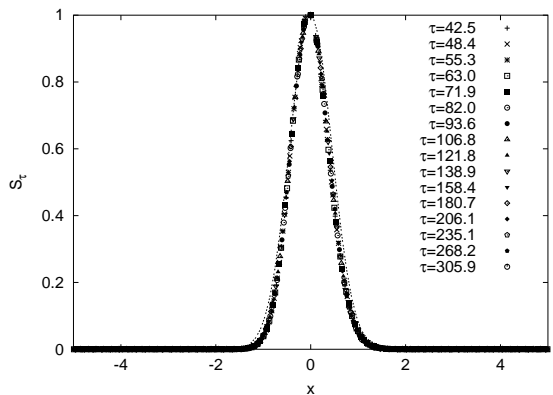}}}
\caption{The $S_\tau(x)$ scaled function of $\Pi_\tau(x)$ for different
$\tau$ values with field ${\bf E}=(0.3,0.48)$ 
(the same configuration as for Fig.~\ref{histfig}). 
The dashed curve shows the Gaussian function $S_G(x)=e^{-\sqrt{2\pi}x^2}$.}
\label{scalfig}
\end{figure}
Observing in Fig.~\ref{mntsfig} that the standard deviation of 
$\Pi_\tau(x)$ is proportional to $1/{\sqrt{\tau}}$ 
(thus $\Pi^{\left(max\right)}_\tau \sim \sqrt{\tau}$), 
one can see that Fig.~\ref{scalfig} shows $\pi_\tau$ fluctuations 
of $O(1/{\sqrt{\tau}})$. 
It is worth noting that fluctuations of this magnitude also constitute
the range where the central limit theorem (CLT) could in principle guarantee 
the Gaussian form of $S_\tau(x)$ for high enough $\tau$ values. 
Indeed, by examining Fig.~\ref{scalfig} we can conclude that for high 
$\tau$ values {\em all} the measured points of $S_\tau(x)$ 
are close to the Gaussian $S_G(x)=e^{-\sqrt{2\pi}x^2}$.

\subsection{The moments}
\begin{figure}
\scalebox{0.3}{\rotatebox{-90}{\includegraphics{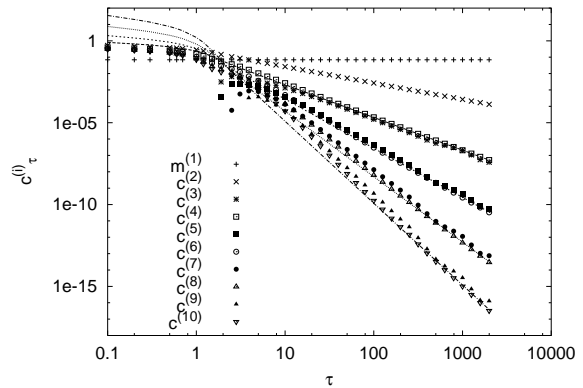}}}
\caption{The central moments $c^{(i)}_{\tau}$ of $\Pi_\tau(x)$ in the 2D 
periodic LG.
In this configuration ${\bf E}=(0.3,0.48)$, the collision number 
is $1.5\times10^{7}$.
The continuous lines show the behavior expected from Gaussian moments: 
$c^{(4)}_G=3\,{c^{(2)}}^2$, $c^{(6)}_G=15\,{c^{(2)}}^3$, $c^{(8)}_G=105\,{c^{(2)}}^4$, 
$c^{(10)}_{G}=945\,{c^{(2)}}^5$, etc., calculated from $c^{(2)}$.}
\label{mntsfig}
\end{figure}

In order to characterize the deviation of the PDF from a Gaussian,
we compute its moments and cumulants from the simulated distribution. 
Let $m^{(i)}$, $c^{(i)}$ and $q^{(i)}$ denote the $i$'th order moment, 
central moment and cumulant of $\Pi_\tau(x)$ respectively
($i=1,2,3,\ldots$) \cite{moments}. 
First we deal with the central moments of $\Pi_\tau(x)$ for a given 
configuration of the periodic LG, shown in Fig.~\ref{mntsfig}.
Based on this figure, we can make several important observations.
\begin{enumerate}
\item 
The central moments of order $i$  $c^{(i)}_\tau$  exhibit
power law dependence on $\tau$ with exponents $\beta_i$ depending on $i$: 
$c^{(i)}_\tau \approx \alpha_i \tau^{-\beta_i}$;
Table~(\ref{exptbl}) lists the numerical values of $\alpha_i$ and 
$\beta_i$ found in the simulation.
\item 
The exponents $\beta_i$ are found to form pairs, at least approximately: 
$\beta_3 \approx \beta_4$, $\beta_5 \approx \beta_6$, and so on.
\item
The even order central moments behave the same way as for a Gaussian 
distribution, i.e., they all can be computed from the variance
alone. 
The Gaussian moments $c_{G}^{(i)}$ calculated in this way from $c^{(2)}$ 
are shown as
continuous lines in the figure, and indeed, they seem to run very 
close to the actual measured values of the even order central moments.
\end{enumerate}
Analyzing several configurations with various strengths and directions
of the external field, we found that the $\beta_i$ exponents are
approximately the same in all the configurations.
In contrast, the $\alpha_i$ factors do depend on the details of the
chosen configuration.
\begin{table}
    \begin{tabular}[t]{c||c|c}
      $i$ & $\beta_i$ & $\alpha_i$\\
      \hline\hline \\
      $2$   & $0.961 \pm 0.003 \approx 1$ & $0.224 \pm 0.003$ \\
      $3$   & $2.047 \pm 0.016 \approx 2$ & $0.225 \pm 0.017$ \\
      $4$   & $1.965 \pm 0.005 \approx 2$ & $0.190 \pm 0.005$ \\
      $5$   & $3.046 \pm 0.048 \approx 3$ & $0.614 \pm 0.048$ \\
      $6$   & $3.005 \pm 0.004 \approx 3$ & $0.330 \pm 0.007$  \\
      $7$   & $4.023 \pm 0.024 \approx 4$ & $1.616 \pm 0.181$  \\
      $8$   & $4.052 \pm 0.005 \approx 4$ & $0.868 \pm 0.021$  \\
      $9$   & $4.937 \pm 0.038 \approx 5$ & $3.742 \pm 0.656$  \\
      $10$  & $5.068 \pm 0.187 \approx 5$ & $2.631 \pm 0.187$  \\
    \end{tabular}
\caption{The exponent $\beta_i$ and the prefactor $\alpha_i$ of the $i$-th 
central moment shown on Fig.~\ref{mntsfig}. 
These values are computed using the NLLS Marquardt-Levenberg 
algorithm included in {\tt gnuplot} to fit the measured data points
with a power function on the interval $[100,1000]$. 
The errors presented here come from the fitting method.
The odd order moments seem to run close to their even successors:
$\beta_{2n+1} \approx \beta_{2(n+1)} \approx n+1$. }
\label{exptbl}
\end{table}

\begin{figure}
\scalebox{0.3}{\rotatebox{-90}{\includegraphics{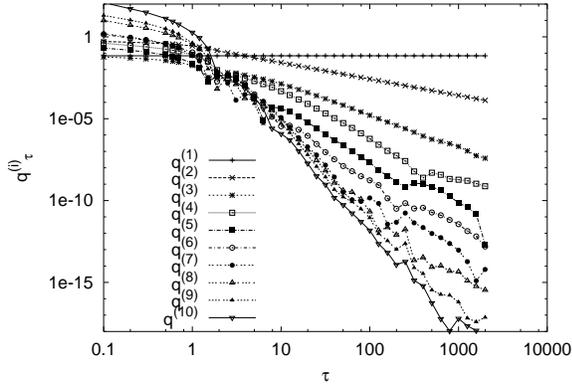}}}
\caption{The absolute values of the cumulants $q_\tau^{(i)}$ of $\Pi_\tau(x)$ 
in the periodic LG.
The configuration is the same as for Fig.~\ref{mntsfig}.
$\Pi_\tau(x)$ converges to a Gaussian in the $\tau \to \infty$ limit. 
Since $q^{(3)} \sim \tau^{-2}$, the convergence of $\Pi_\tau(x)$ to a Gaussian 
is power like; the deviation is on the order of $\tau^{-2}$.}
\label{cmsfig}
\end{figure}

Now we compare our findings to the expectations based on a Gaussianity 
assumption. Because of the power law dependence in the odd order central 
moments, we can say that $\Pi_\tau(x)$ is definitely {\em not} Gaussian 
for any finite value of $\tau$ (they all should be zero for a Gaussian 
distribution).
On the other hand, the distributions $\Pi_\tau(x)$ do seem to converge 
to a Gaussian in the limit $\tau \to \infty$. 
This can be checked in Fig.~\ref{cmsfig}
showing the behavior of the cumulants: 
all of the higher ($i>2$) order cumulants go to zero faster
than those of order $i=1, 2$ in the $\tau \to \infty$ limit.

Another important consequence of the pairing of the $\beta_i$ exponents is
that  $\Pi_\tau(x)$ can not be scaled in the sense of Eq.~(\ref{eqscal}).
Indeed, the equation $S_{\tau}(x)=S(x)$ results that
$\Pi_{\tau}\left(x+m\right)= 
\Pi^{\left(max\right)}_\tau S\left( \Pi^{\left(max\right)}_\tau x \right)$,
since $x_{max}=m$ for high $\tau$ values. 
Substituting this equation into the definition of the central moments 
and performing the variable transformation 
$y=\Pi^{\left(max\right)}_\tau x$ yields
\begin{eqnarray}
c_i=\int_{-\infty}^{\infty}\,\mathrm{d}x \Pi_\tau(x+m) x^i= \nonumber \\
\int_{-\infty}^{\infty}\,\mathrm{d}x\Pi^{\left(max\right)}_\tau 
S\left( \Pi^{\left(max\right)}_\tau x \right) x^i=\\
\left(\Pi^{\left(max\right)}_\tau\right)^{-i} 
\int_{-\infty}^{\infty}\,\mathrm{d}y S(y) y^i \nonumber\,.
\end{eqnarray}
The last expression shows that if $S_{\tau}(x)=S(x)$ were true in this
system, then the $\beta_i$ exponents should obey the $\beta_i=\frac{i}{2}$
identity, since $\Pi^{\left(max\right)}_\tau \sim \sqrt{\tau}$ for high $\tau$
values.  Table~\ref{exptbl} shows that the odd order $\beta_i$ exponents
deviate significantly from $\frac{i}{2}$.  This behavior indicates
that the scaling property suggested by Fig.~{\ref{scalfig}} can only
be treated as a good approximation and not as an exact result in the
GIK thermostated LG.  
This also raises the question whether similar claims of scaling, 
based on the visual properties of the PDF in other systems as e.g.\ 
in Ref.~\cite{observ}, could also be supported by the behavior of the
moments.

\subsection{The Fluctuation Formula} 
\begin{figure}
\scalebox{0.3}{\rotatebox{-90}{\includegraphics{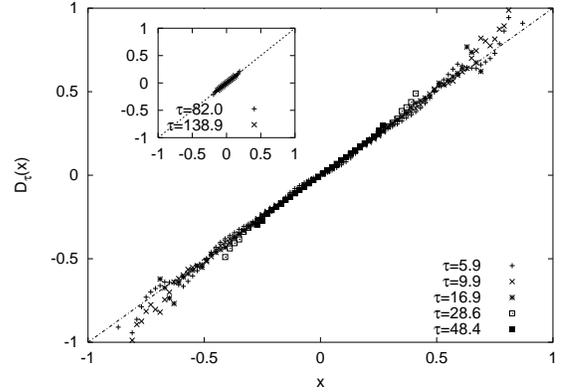}}}
\caption{$D_\tau(x)$ in the same configuration as for Fig.~\ref{histfig}. 
The inset shows $D_\tau(x)$ for higher $\tau$ values. 
The inset and the figure have the same axes.}
\label{fluctfig}
\end{figure}

We can rewrite the FF in the terms of $\Pi_\tau(x)$ as
\begin{eqnarray}
  \label{flformPi}
  \lim_{\tau\to\infty}
  \frac{1}{E}\frac{1}{\tau}\,\ln \frac{\Pi_\tau(x)}{\Pi_\tau(-x)}=x\,. 
\end{eqnarray}
In order to visualize the FF, we introduce the quantity
\begin{eqnarray}
\label{Dtau}
    D_\tau(x)=\frac{1}{E}\frac{1}{\tau}\,
         \ln \frac{\Pi_\tau(x)}{\Pi_\tau(-x)}.
\end{eqnarray}  
If the FF is satisfied in the system,
then $D_\tau(x)$ is linear with slope 1 in the $\tau \to \infty$ limit.
Analyzing the behavior of this quantity in Fig.~\ref{fluctfig}
one can make two important observations:
\begin{enumerate}
\item The FF is satisfied quite well for both {\em low} $\tau$ values and for 
  large ones (in this context, low $\tau$ value means 6-8 times 
  the average time between two subsequent collisions).

\item As $\tau$ gets higher, the interval where $D_\tau(x)$ can be plotted is getting 
  narrower and narrower. This is a consequence of the fact that 
  for higher $\tau$ values large fluctuations of $\xi_\tau$ 
  become less and less probable. In other words fluctuations 
  outside the plotted range are practically {\em unobservable}.
\end{enumerate}  
We note that we tested several other configurations
with various choices of the magnitude and direction of the electric  
field ${\bf E}$ in the ergodic range \cite{ergodic},
and we didn't find any relevant deviation from the behavior
presented above. 

\subsection{The Gaussian approximation}
Since we have seen that the even order central moments behave 
in a Gaussian way, it is worth checking if the subcondition that leads 
to the FF for Gaussian distributions holds for $\Pi_\tau(x)$.
This subcondition can be derived by substituting  
$\Pi_\tau(x)=\frac{1}{\sqrt{2 \pi} \sigma_\tau}
 \exp {\left( \frac{-(x-m_\tau)^2}{2 \sigma_\tau^2} \right) }$ 
into Eq.~(\ref{flformPi}). 
This leads to the following formula:
\begin{eqnarray}
\label{flgauss}
\lim_{\tau \to \infty} 
  \frac{1}{E} \frac{1}{\tau} \frac{2 m_\tau} {\sigma_\tau^2}=1\,,
\end{eqnarray}
where $m_{\tau}=m^{(1)}_\tau$ and $\sigma_\tau^2=c^{(2)}_\tau$
according to the previous notation.

\begin{figure}
\scalebox{0.3}{\rotatebox{-90}{\includegraphics{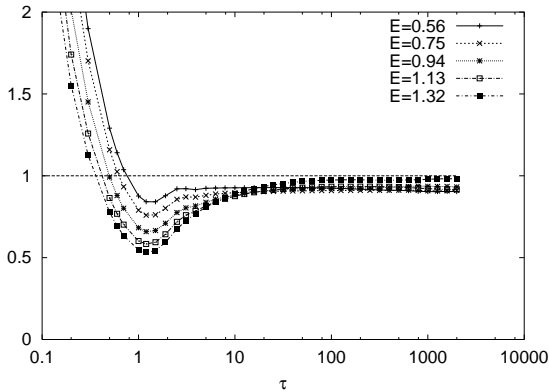}}}
\caption{The quantity $\frac{1}{E} \frac{1}{\tau} \frac{2 m_\tau}{\sigma_\tau^2}$ at different field strengths ${\bf E}$ in the 2D periodic Lorentz Gas. 
In these configurations ${\bf E}$ is parallel with $(0.3,0.48)$ 
and the collision number is $\approx 1.5 \times 10^7$.
The curves deviate from 1 for all the $\tau$ values presented on the figure,
however seem to converge to 1 in the $\tau \to \infty$ limit. }
\label{gauss}
\end{figure}

In Fig.~\ref{gauss} we can verify that the 
$\frac{1}{E} \frac{1}{\tau} \frac{2 m_\tau} {\sigma_\tau^2}=1$
identity is violated for all finite $\tau$ values but satisfied
in the $\tau \to \infty$ limit.
On the other hand, we have seen in Fig.~\ref{fluctfig} 
that the FF is valid with reasonable precision even for low values of $\tau$. 
These observations indicate that:
\begin{enumerate}

\item For finite $\tau$ values the the odd order moments also 
contribute to the fulfillment of the FF, thus the Gaussian approximation of 
the PDF $\Pi_\tau(x)$ violates the FF. This behavior is also supported
by the analysis performed in Ref.~\cite{exptest}. 

\item For high $\tau$ values the observable part of the PDF 
$\Pi_\tau(x)$ seems to converge to a Gaussian (see. Fig.~{\ref{scalfig}})
and the Gaussian approximation of the PDF $\Pi_\tau(x)$ satisfies the FF.

\end{enumerate}
This is in agreement with the observed behavior of the cumulants 
indicating a convergence to Gaussian distributions for $\tau \to \infty$;
indeed, for Gaussian distributions the FF is equivalent 
to Eq.~({\ref{flgauss}).

\begin{figure}[ht]
\scalebox{0.3}{\rotatebox{-90}{\includegraphics{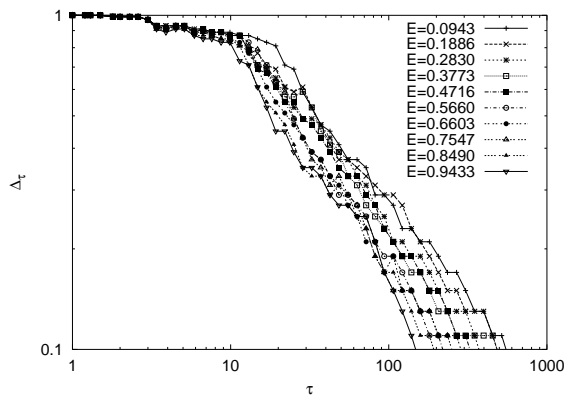}}}
\caption{The dependence of $\Delta_{\tau,E}$ on $\tau$ for different field
strengths. 
The direction of ${\bf E}$ is parallel with $(5,8)$ but its magnitude varies.
The curves appear to be linear in the dominant region on the
log-log plot, suggesting a power law dependence on $\tau$.
}
\label{delta}
\end{figure}
\subsection{The observability of large fluctuations}
Since we have shown that for high $\tau$ values 
neither $\Pi_\tau(x)$ nor $D_\tau(x)$
can be observed on the whole interval $[-1,1]$,
we introduce $\Delta_{\tau,E}$ to quantify where $D_\tau(x)$ can be observed, 
i.e. where the FF can be tested.
Let $[-\Delta_{\tau,E},\Delta_{\tau,E}]$ denote the interval where  $D_\tau(x)$ can be reconstructed 
from a given time series in a specific configuration. 
Data from our numerical simulations are summarized in Fig.~\ref{delta}.
It shows that $\Delta_{\tau,E}$ has a power-law dependence on $\tau$: 
\begin{eqnarray}
 \Delta_{\tau,E} \sim \tau^{-\delta(E)} \,,
\end{eqnarray} 
where the exponent $\delta$ tends to increase with increasing field strengths.

\section{Conclusion}
\label{sec:concl}

In this paper we have investigated numerically the PDF of the EPR fluctuations 
for both low and high averaging time intervals in the 2D periodic LG 
thermostated by the GIK thermostat. 
We have shown that for {\em low} averaging time intervals 
the EPR fluctuations are definitely not Gaussian, 
but they satisfy the Gallavotti-Cohen fluctuation formula.
For {\em high} averaging time intervals the EPR fluctuations are converging 
to a Gaussian distribution and still obeying the FF, 
thus the Gaussian approximation of the EPR fluctuations seems to be 
better and better as $\tau$ gets higher in this model.
We have analyzed the moments of the corresponding PDF to characterize 
its functional form and found a special power-law dependence 
(pairing of the $\beta_i$ exponents) of the central moments on the averaging 
time interval. 
Our results raise the question whether the $\beta_i$-pairing could be observed
in other systems satisfying the FF, or this is just a special property of
the GIK thermostated Lorentz gas.   

\begin{acknowledgments}
The authors are grateful to Tam{\'a}s T{\'e}l for fruitful discussions
and a careful reading of the manuscript. 
Furthermore the first author thanks J{\"u}rgen Vollmer for valuable e-mail 
conversations.
This work was supported by the Bolyai J{\'a}nos Research Grant of
the Hungarian Academy of Sciences and by the Hungarian Scientific 
Research Foundation (Grant No.\ OTKA T032981).
\end{acknowledgments}



\begin{thebibliography}{99}
\bibitem{moments} Obviously the quantity in question is $\Xi_\tau(x)$, 
however it is easier to work with $\Pi_\tau(x)$. 
The connection between the analogous quantities of 
the two PDFs $\Xi_\tau(x)$ and $\Pi_\tau(x)$ can easily be derived.
Let $\tilde{m}^{(i)}$, $\tilde{c}^{(i)}$ and $\tilde{q}^{(i)}$ 
denote the $i$'th order moment, central moment and cumulant pertaining to
$\Xi_\tau(x)$, then:
$\tilde{m}^{(i)}=E^i m^{(i)}$, $\tilde{c}^{(i)}=E^i c^{(i)}$ and 
$\tilde{q}^{(i)}=E^i q^{(i)}$.
\end{thebibliography}
\end{document}